\documentclass[12pt]{article}

\usepackage{scicite}

\usepackage{graphicx}

\usepackage{times}

\newenvironment{sciabstract}{%
\begin{quote} \bf}
{\end{quote}}

\title{Five-dimensional imaging of freezing emulsions with solute effects} 

\author
{Dmytro Dedovets$^1$, C\'{e}cile Monteux,$^2$ Sylvain Deville$^{1\ast}$\\
\\
\normalsize{$^{1}$LSFC, UMR 3080 CNRS/Saint-Gobain CREE, Cavaillon, France}\\
\normalsize{$^{2}$SIMM, UMR 7615 CNRS-ESPCI-Universit\'e Pierre Et Marie Curie, ESPCI, Paris, France}\\
\\
\normalsize{$^\ast$To whom correspondence should be addressed; E-mail:  sylvain.deville@saint-gobain.com.}
}

\date{}

\begin{document} 


\baselineskip24pt


\maketitle 


\begin{sciabstract}
The interaction of objects with a moving solidification front is a common feature of many industrial and natural processes such as metal processing, the growth of single-crystals, the cryopreservation of cells, or the formation of sea ice. Solidification fronts interact with objects with different outcomes, from the total rejection to their complete engulfment. We image the freezing of emulsions in 5D (space, time, and solute concentration) with confocal microscopy. We show the solute induces long-range interactions that determine the solidification microstructure. The local increase of solute concentration enhances premelting, which controls the engulfment of droplets by the front and the evolution of grain boundaries. Freezing emulsions may be a good analogue of many solidification systems where objects interact with a solidification interface.
\end{sciabstract}

Solidification fronts interact with hard or soft objects such as particles, bubbles, droplets, or cells, in three different ways: objects can be engulfed, rejected for some time and then encapsulated, or rejected for an extended period of time~\cite{Asthana1993}. Different outcomes are desired, depending on the context. Particle-reinforced metal alloys require homogeneous particles distribution in the matrix and immediate engulfment is therefore preferred~\cite{Zhang2006}. In contrast, for single-crystal growth, a complete rejection of impurities is essential~\cite{Bunoiu2010}. The rejection of dissolved hydrogen can form gas bubbles which are eventually trapped by solidification fronts, becoming a major source of defect in many metals and alloys~\cite{Gupta1992}. Reproductive and red blood cells can be physically damaged when engulfed between growing ice crystals during cryopreservation procedures~\cite{Korber1988,Scott2005}. The processing of porous materials by ice-templating relies on the complete segregation of particles by the solidification front~\cite{Deville2017i}. Our ability to control the solidification microstructure is essential in all these domains.

The dynamics of interfaces and objects are an inherent feature of these phenomena, but are extremely difficult to observe because of the time-scale (solidification fronts moving rapidly), space-scale (submicronic objects), and sometimes high-temperature. In addition, the role of solute is often important in these systems, from the gas and impurities in metals to the additives used in cryopreservation procedures or food engineering. Being able to follow in situ the concentration of the various phases in presence of objects is crucial and yet extremely challenging. Direct photometric measurements of the solute concentration in thin layers of solution during solidification have been reported~\cite{Diller1990}, but only in 2D, and not in presence of particles. Confocal Raman microspectroscopy was used to measure the static concentration of a cryoprotective agent~\cite{Dong2010} in frozen cells encapsulated in an ice matrix.

Progress in solidification research has thus largely relied on the studies of analogues: transparent materials that solidify near room temperature, using optical microscopy~\cite{Farup2001situ}. However, 3D imaging cannot be performed and chemistry cannot be resolved. Thanks to recent progresses in X-ray computed tomography, we can obtain 3D, time-lapse images of the microstructure evolution in situ during the solidification of metals~\cite{Tolnai2012}. The volume investigated is nevertheless limited, particles can barely be resolved, the temporal resolution is not good enough to capture the interaction events~\cite{Daudin2015}, and quantifying the solute concentration in the presence of objects is still not possible. It is thus fundamentally difficult to track in situ the development of 3D solidification microstructures where objects interact with a solid/liquid interface~\cite{Deville2017c}, even more so in presence of solute effects. 

We studied the solidification process by confocal microscopy, translating a Hele-Shaw cell filled with an oil-in-water emulsions at constant velocity $V$ through a fixed temperature gradient $\Delta T$ (Fig.~1A and movie~S1). This way the solidification front propagates with a velocity $V$ through the sample. A typical confocal image of the freezing emulsion is shown in Fig.~1B. The sulforhodamine B fluorescence served as a local proxy for the surfactant concentration (Fig.~S1).

We used the droplets trajectories to measure their velocity in the observation frame, $U'$ and deduce their velocity in the sample frame, $U$=$V$-$U'$ (Fig.~S2). An example of the variation of the droplet velocity $U$ with time (Fig.~2A) shows that it goes through a maximum as the droplet approaches the solidification front. The droplets start being repelled from the front at a distance $L_v$ of the order of $100\mu m$, accelerate up to a maximal velocity $U_{max}$ of $1.2\mu m/s$, and rapidly decelerate as they cross the solidification front (Fig.~2B). The droplets are eventually completely engulfed and stop moving. The time during which the droplets are repelled by the front before being engulfed is called the interaction time, plotted as a function of interface velocity in Fig.~2D. During the interaction time, droplets are displaced by a large distance, from $2\mu m$ to more than $100\mu m$ (Fig.~2E).

To understand the origin of the droplet displacement, we measure the solute concentration profile as a function of the distance from the solidification front. The solute concentration (Fig.~2C) is constant far from the front and then progressively increases. The concentration abruptly drops as we cross the interface because of the low solubility of solute in ice. The droplet starts being repelled by the interface at a distance $L_v$ close to $L_c$--the distance at which the concentration of solute increases. This is more apparent when we plot $L_v$ as a function of $L_c$ for several front velocities (Fig.~2F), which shows that $L_v$ increases linearly with $L_c$.

The presence of a surfactant in the system results in driving forces that can displace droplets over up to hundreds of microns from the solidification front, on time-scale of tens to hundreds of seconds. These results depart from the behavior observed and predicted by most physical models of the interaction of a particle with a solidification front~\cite{Asthana1993} which, in absence of solute, assumed a dominant role of the thermomolecular forces in the interactions between the object and the front~\cite{Korber1988}. However, such interactions are only effective on very short distances (nanometers). The distances measured here are several orders of magnitude larger. In the presence of a solute such as a surfactant, the thermomolecular forces seem to play a minor role. 

We suggest that the driving force for the motion of the droplets is diffusiophoresis or solutal Marangoni effect, which are known to induce micron scale motion of droplets or particles in gradients of solute concentration ~\cite{Anderson1982,Weber2017}. The droplet velocity increases with the gradient of surfactant concentration along the $x$ axis (Fig.~2G). Diffusiophoresis is due to a gradient of osmotic pressure, while Marangoni flow is due to a gradient of surface tension along the droplet surface. In our case, the bulk surfactant concentration is two orders of magnitude above the critical micelle concentration (CMC) and rises by a factor of 3 to 10 near the solidification front. Therefore, we expect a weak variation of the surfactant monomer concentration and of the surface tension along the $x$ axis but a significant variation of the concentration of micelles, which may lead to a diffusiophoretic motion.

The long range interactions between the droplets and the front have a strong influence on the solidification microstructure. When the droplets displacement (Fig. 2E) is larger than the inter-droplet distance, the droplets form clusters as they get engulfed by the front (top of Fig. 2H, obtained for a front velocity of $1\mu m/s$). At $2\mu m/s$, the interaction time is shorter and almost no aggregates are formed. For a front velocity of $3\mu m/s$, the droplets displacement is significantly smaller than mean inter-droplet distance, which results in the preservation of the original spatial distribution of droplets upon solidification. In all these cases, the droplets are engulfed by the growing ice, but the microstructure of the solidified sample differs, depending on the solidification front velocity.

The majority of physical models developed for the past 60 years to describe the interaction of particles with a solidification front focused on the criterion of critical particle size for a given velocity (or conversely the critical velocity for a given size). The results shown here demonstrate how important the dynamics of interactions with the solidification front can be. In situations where the distribution of objects should be retained in the solidified microstructure, such as particle reinforcements in a metal matrix, the displacement of objects by the solidification front is a much more relevant parameter to understand and control the solidification microstructures. 

We investigated the solute redistribution in the premelted film at the ice grains boundaries and around the droplets after their engulfment. Premelting describes the existence of liquid films at solid surfaces at temperatures below the bulk freezing temperature, a phenomenon common to many solids, including ice. Premelting has consequences in a wide range of biological, geophysical, and technological processes~\cite{Wettlaufer2006}, such as the heaving of frozen ground~\cite{Dash1999}, glacier motion, weathering, material transport through ice, using ice core as climate proxies~\cite{Rempel2010}, or the mobility of particles in solid materials. The driving force for melting is a reduction of interfacial energy between the particles and the matrix. The concentration of impurities or solute that depress the freezing point of the melt can greatly enhance the interfacial premelting, which is then called solute premelting. Because of their small dimensions, imaging premelted films remains a challenge.

Confocal microscopy allows in situ, 3D imaging of the solute redistribution in the solidifying sample with a semi-quantitative estimation of its concentration. In the solidified part of the sample, the solute is concentrated in the premelted film at the ice crystals grain boundaries (Fig.~3A, region 1) and around the engulfed oil droplets (Fig.~3A, region 2). Because the concentrated solute broadens the thickness of these wetting films, they can be observed.

At low solidification front velocities (1 and 2 $\mu m/s$), the interface is planar. Increase of the interface velocity to 3 $\mu m/s$ (Fig.~3B) results in the formation of grain boundaries--periodical liquid films of concentrated solute. Their number and spacing increase with the interface velocity (Fig.~S1). 

The lateral growth of ice crystals results in a progressive thinning of the premelted films at the grain boundaries, eventually leading their breakup into spherical droplets by an interfacial instability (Fig.~3B). These microstructures are analogue to those developed in metal alloys with a miscibility gap~\cite{Majumdar1996} (Fig.~3C, D). The transition from planar to a cellular front and the destabilization of the premelted film into droplets depend on both the temperature gradient and the solidification velocity (Fig.~3E), similar to that in metals~\cite{Ratke2005}. 

We can capture the instability initiation and propagation dynamics (Fig.~3F). The growing geometrical instabilities results in the formation of side-branches that elongate and break up into droplets. This process is quite slow, on the time-scale of tens of seconds, which is a result of low interfacial tension between the solid and liquid phase. The development of these instabilities in 3D (Fig.~S4) reveal that they initiate where the liquid film is the thinnest. This instability is analogous to the pinch off in dendrites observed in metal alloy solidification, which is driven by capillarity~\cite{Neumann-Heyme2015}.

The premelted film also plays a critical role in the droplet engulfment by the solidification front (Fig.~4A). At the approach of the droplet, the solidification front bends away from it due to the formation of a pocket of concentrated solute which colligatively depresses the freezing point of water. Analytical models predict a deflection of the solidification front towards an object if the thermal conductivity of the latter is lower than that of the melt~\cite{Park2006}. The thermal conductivity of oil (propyl benzoate) is smaller than that of water ($0.141 Wm^{-1}K^{-1}$ versus $0.569 Wm^{-1}K^{-1}$) and thus the formation of a convex interface is expected, which is what we observe in the absence of solute (Fig.~S5). However, in the presence of solute, we observe only concave interfaces. The solute is thus able to reverse the bending of the front, which indicates the dominant role of the solute in droplet/interface interaction. 

The ice continues to grow over and around the droplet, until the latter is completely engulfed. 3D imaging (Fig.~4B) provides a complete picture of the morphological evolution of the premelted film around the droplet. As the solidification proceeds, the solute-rich liquid pocket, initially located in front of the oil droplet, passes on the opposite side of it--away from growing ice. From the fluorescence intensity, we obtain semi-quantitative information of the evolution of the solute concentration in the premelted film around the oil droplet (Fig.~4C and S6). We then correlate the solute concentration with the thickness of the liquid film (Fig.~4D), to reveal the dynamics of the process, and see how the premelted film evolves towards its equilibrium thickness ($5\mu m$ in this case). 

For the droplet size investigated, comet tail-shaped segregation pattern were systematically observed (Fig.~4E). These observations may be a good analogue for industrial processes where pores and particles can have an impact on the segregation pattern (Fig.~4F--H). Understanding the solute redistribution mechanisms could thus be an important step towards a better control of the solidification microstructures. 

Solute also plays a critical role in the cryopreservation of red blood cells, platelets, fibroblasts, and reproductive cells~\cite{Hubel1997}. Cryoprotection agents are now systematically used to improve the survival rate. The crystal growth pattern and concentration of these agents between the growing crystals, along with the cells, impact their survival rates~\cite{Korber1988,Ishiguro1994,Uemura2015}. Quantitative studies of solute concentration during freezing (but without cells) were done in thin films, by photometric measurements~\cite{Diller1990}. However, when the integration of concentration is done through the depth of the sample, the peak solute concentration may be underestimated if the solidification microstructure is not homogeneous. Our results show that, because of the complex solute concentration and redistribution pathways, 5D imaging is required to better estimate the solute location and concentration. Our approach may help better understand the physical damages inflicted to cells during the cryopreservation procedures and optimize such procedures.

The features observed during the freezing of emulsions seems thus to be good analogues of many solidification systems. The rapid 3D confocal microscopy imaging, along with the ability to map the solute concentration, provides new insights into the dynamics and mechanisms of solidification in the presence of foreign hard (particles) or soft (droplets, bubbles, cells) objects. The approach proposed here is versatile and can be easily adapted to a variety of solidification systems, from geophysics to food engineering or biology. We expect confocal microscopy to be a welcome addition to the toolbox available for the studies of interfacial evolution problems in three dimensions~\cite{Rowenhorst2012}.

\bibliography{scibib}

\begin{thebibliography}{10}

\bibitem{Asthana1993}
R.~Asthana, S.~N. Tewari, {\it J. Mater. Sci.\/} {\bf 28}, 5414 (1993).

\bibitem{Zhang2006}
L.-F. Zhang, {\it Journal of Iron and Steel Research International\/} {\bf 13},
  1 (2006).

\bibitem{Bunoiu2010}
O.~M. Bunoiu, T.~Duffar, I.~Nicoara, {\it Prog. Cryst. Growth Charact.
  Mater.\/} {\bf 56}, 123 (2010).

\bibitem{Gupta1992}
A.~K. Gupta, B.~K. Saxena, S.~N. Tiwari, S.~L. Malhotra, {\it J. Mater. Sci.\/}
  {\bf 27}, 853 (1992).

\bibitem{Korber1988}
C.~K{\"{o}}rber, {\it Q. Rev. Biophys.\/} {\bf 21}, 229 (1988).

\bibitem{Scott2005}
K.~L. Scott, J.~Lecak, J.~P. Acker, {\it Transfus. Med. Rev.\/} {\bf 19}, 127
  (2005).

\bibitem{Deville2017i}
S.~Deville, {\it Scr. Mater.\/} pp. 1--6 (2017).

\bibitem{Diller1990}
S.~Kourosh, K.~R. Diller, M.~E. Crawford, {\it Int. J. Heat Mass Transf.\/}
  {\bf 33}, 39 (1990).

\bibitem{Dong2010}
J.~Dong, J.~Malsam, J.~C. Bischof, A.~Hubel, A.~Aksan, {\it Biophys. J.\/} {\bf
  99}, 2453 (2010).

\bibitem{Farup2001situ}
I.~Farup, J.-M. Drezet, M.~Rappaz, {\it Acta materialia\/} {\bf 49}, 1261
  (2001).

\bibitem{Tolnai2012}
D.~Tolnai, {\it et~al.\/}, {\it Acta Mater.\/} {\bf 60}, 2568 (2012).

\bibitem{Daudin2015}
R.~Daudin, S.~Terzi, P.~Lhuissier, L.~Salvo, E.~Boller, {\it Mater. Des.\/}
  {\bf 87}, 313 (2015).

\bibitem{Deville2017c}
S.~Deville, {\it Freez. Colloids Obs. Princ. Control. Use\/} (Springer
  International Publishing, 2017), pp. 47--90.

\bibitem{Anderson1982}
J.~L. Anderson, M.~E. Lowell, D.~C. Prieve, {\it J. Fluid Mech.\/} {\bf 117},
  107 (1982).

\bibitem{Weber2017}
C.~A. Weber, C.~F. Lee, F.~J{\"{u}}licher, {\it New J. Phys.\/} {\bf 19},
  053021 (2017).

\bibitem{Wettlaufer2006}
J.~S. Wettlaufer, M.~G. Worster, {\it Annual Review of Fluid Mechanics\/} {\bf
  38}, 427 (2006).

\bibitem{Dash1999}
J.~G. Dash, H.~Fu, J.~S. Wettlaufer, {\it Reports on Progress in Physics\/}
  {\bf 58}, 115 (1999).

\bibitem{Rempel2010}
A.~W. Rempel, {\it J. Glaciol.\/} {\bf 56}, 1122 (2010).

\bibitem{Majumdar1996}
B.~Majumdar, K.~Chattopadhyay, {\it Metallurgical and Materials Transactions
  A\/} {\bf 27}, 2053 (1996).

\bibitem{Ratke2005}
L.~Ratke, {\it Materials Science and Engineering A\/} {\bf 413-414}, 504
  (2005).

\bibitem{Neumann-Heyme2015}
H.~Neumann-Heyme, K.~Eckert, C.~Beckermann, {\it Phys. Rev. E - Stat.
  Nonlinear, Soft Matter Phys.\/} {\bf 92}, 1 (2015).

\bibitem{Park2006}
M.~S. Park, A.~A. Golovin, S.~H. Davis, {\it Journal of Fluid Mechanics\/} {\bf
  560}, 415 (2006).

\bibitem{Hubel1997}
A.~Hubel, {\it Transfus. Med. Rev.\/} {\bf 11}, 224 (1997).

\bibitem{Ishiguro1994}
H.~Ishiguro, B.~Rubinsky, {\it Cryobiology\/} {\bf 31}, 483 (1994).

\bibitem{Uemura2015}
M.~Uemura, H.~Ishiguro, {\it Cryobiology\/} {\bf 70}, 122 (2015).

\bibitem{Rowenhorst2012}
D.~Rowenhorst, P.~Voorhees, {\it Annu. Rev. Mater. Res.\/} {\bf 42}, 105
  (2012).

\bibitem{Ratke2006}
L.~Ratke, A.~Muller, {\it Scr. Mater.\/} {\bf 54}, 1217 (2006).

\bibitem{Catalina2004}
A.~V. Catalina, S.~Sen, D.~M. Stefanescu, W.~F. Kaukler, {\it Metall. Mater.
  Trans. A\/} {\bf 35}, 1525 (2004).

\bibitem{Bystricky1997}
P.~Bystricky, {Plasticity of metal matrix composites reinforced with continuous
  alumina fibres}, Ph.D. thesis, Massachusetts Institute of Technology (1997).

\bibitem{Marcellini2016}
M.~Marcellini, C.~Noirjean, D.~Dedovets, J.~Maria, S.~Deville, {\it ACS
  Omega\/} {\bf 1}, 1019 (2016).

\bibitem{Schindelin2012a}
J.~Schindelin, {\it et~al.\/}, {\it Nat. Methods\/} {\bf 9}, 676 (2012).

\end{thebibliography}

\bibliographystyle{Science}

\section*{Acknowledgements:}
We acknowledge the help of Vittorio Saggiomo (@V\_Saggiomo) for his advices on microfluidics, and F\'elix Ginot for his assistance for some of the Python coding.
\textbf{Funding:} The research leading to these results has received funding from the European Research Council under the European Community's Seventh Framework Programme (FP7/2007-2013) Grant Agreement no. 278004. \textbf{Author contributions:} S.D. designed and supervised the project, S.D., D.D. and C.M. designed the experiments, D.D. designed and built the cooling stage, prepared the emulsions, and carried out the confocal microscopy, D.D. and S.D. analysed the data. All authors discussed the results and implications and wrote the manuscript. \textbf{Competing interest:} The authors declare no competing interest. \textbf{Data and materials availability}: Additional data and code related to this paper are available on Figshare at https://doi.org/10.6084/m9.figshare.5746740.

\begin{figure}
\centering
\includegraphics[width=12cm]{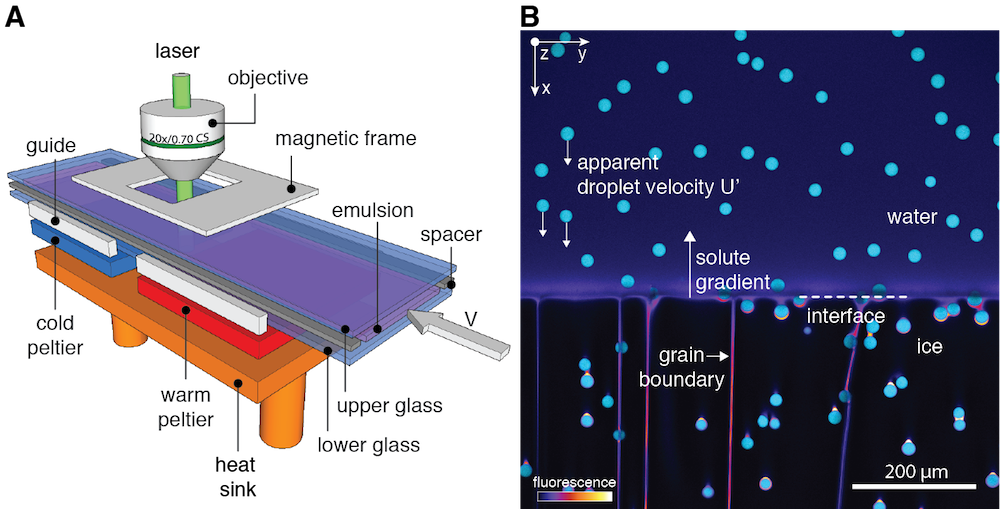}
\caption{Cooling stage and typical confocal image of a freezing emulsion. A: Stage for the directional solidification experiment with a Hele-Shaw cell, placed under the objective of the confocal microscope. B: Typical confocal image of the emulsion during freezing (front velocity of $3\mu m/s$, temperature gradient of $5^\circ C/mm$). The ice is shown in black, the liquid in magenta, and the oil droplets in cyan. \label{fig:figure1}}
\end{figure}

\begin{figure}
\centering
\includegraphics[width=18.3cm]{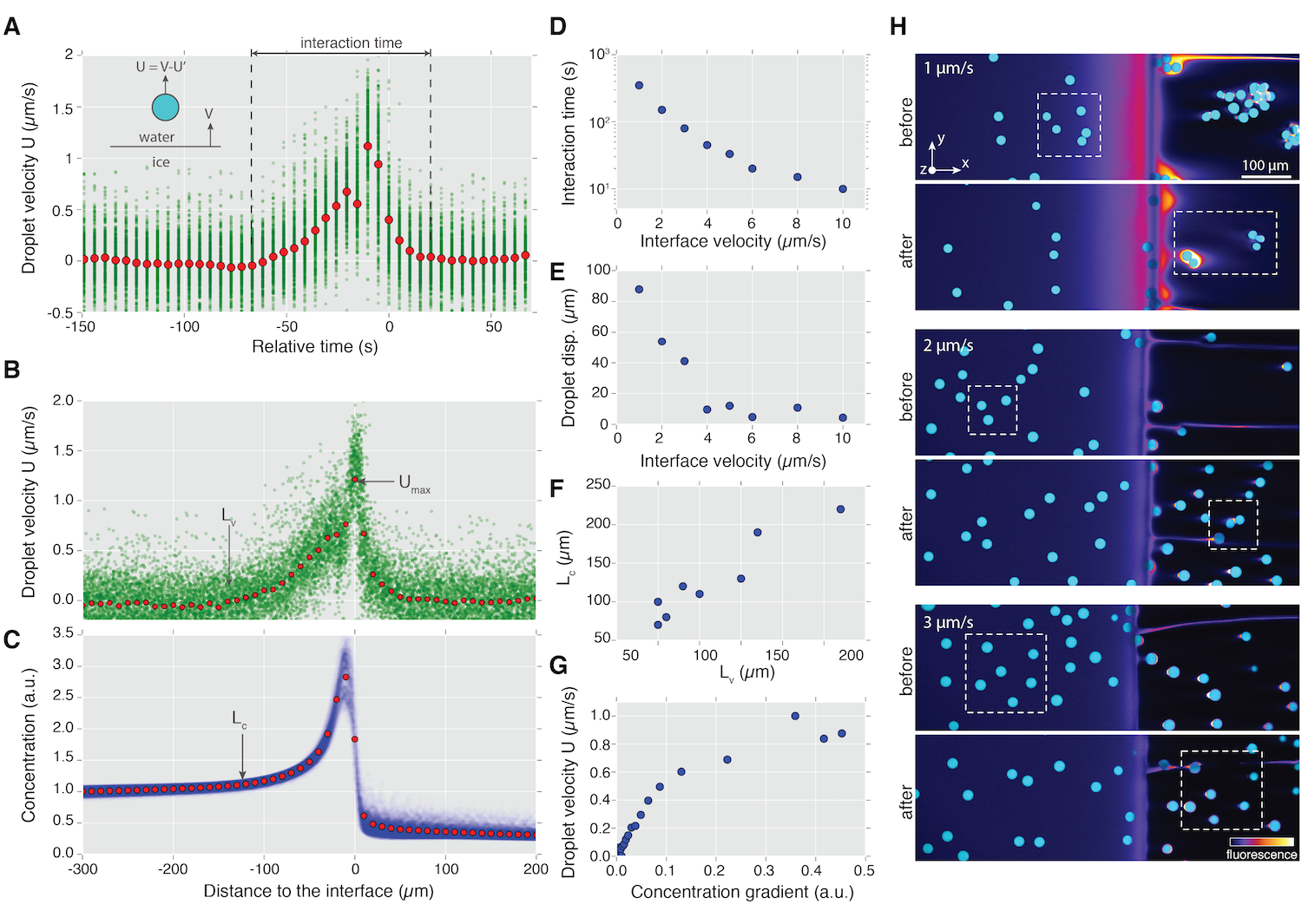}
\caption{Droplets dynamics and solidification microstructures. A: Droplet velocity vs. relative time. Time zero is when the front edge of the droplet is at the solidification front position. B, C: Droplet velocity (B) and solute concentration (C) vs. distance to interface, for a solidification front velocity of $3 \mu m/s$. D, E: Interaction time (D) and droplet displacement (E) vs interface velocity. The error bars do not exceed the size of the circles. F: $L_c$ vs. $L_v$. G: Droplet velocity vs. concentration gradient of solute, for a solidification front velocity of $3 \mu m/s$. H: Impact of the droplet dynamics and interaction time on the development of solidification microstructure (temperature gradient of $5^\circ C/mm$). The dashed line indicates the same groups of droplets before and after passing the interface. \label{fig:figure2}}
\end{figure}

\begin{figure}
\centering
\includegraphics[width=18.3cm]{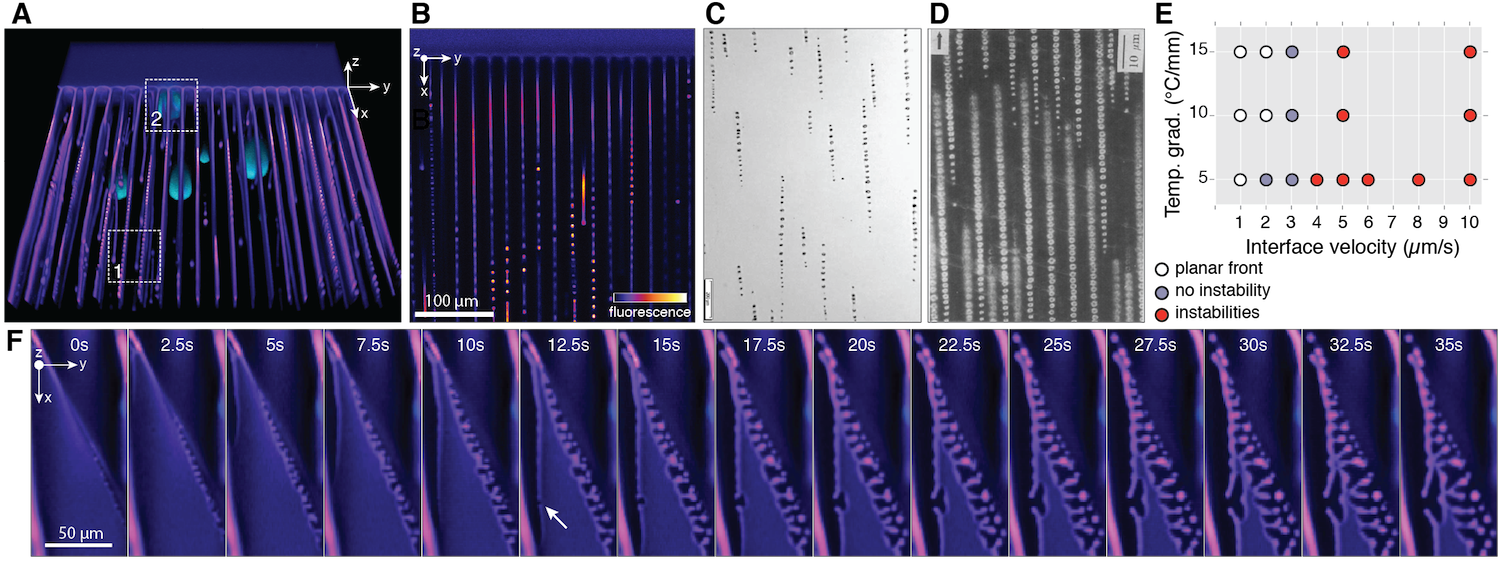}
\caption{Solute premelting around the droplet and at the grain boundaries. A: Typical 3D view of the emulsion during freezing. Premelting plays a role on the destabilization of thin liquid film (concentrated solute) between ice crystals (grain boundaries, region 1), and on the evolution of the droplet/ice interface (region 2). B: String-of-pearls microstructure (solidification front velocity of $3\mu m/s$, temperature gradient of $15^\circ C/mm$) similar to that developed in zinc-bismuth~\cite{Ratke2006} (C) and aluminium-bismuth alloys~\cite{Majumdar1996} (D). E: Stability diagram of the solidification front as a function of the temperature gradient and interface velocity. F: Time-lapse sequence of the dynamics of a premelted film instability at a grain boundary during freezing. The white arrow indicates the location where a second instability is initiated. Credits: (C) Reprinted from Scripta Materialia, 54, Ratke, L. \& M\"{u}ller, A., On the destabilisation of fibrous growth in monotectic alloys, 1217--1220, Copyright (2006), with permission from Elsevier (D) Reprinted by permission from Springer Nature: Springer Nature, Metallurgical and Materials Transactions A~\cite{Majumdar1996}, Copyright (1996).\label{fig:figure3}}
\end{figure}

\begin{figure}
\centering
\includegraphics[width=18.3cm]{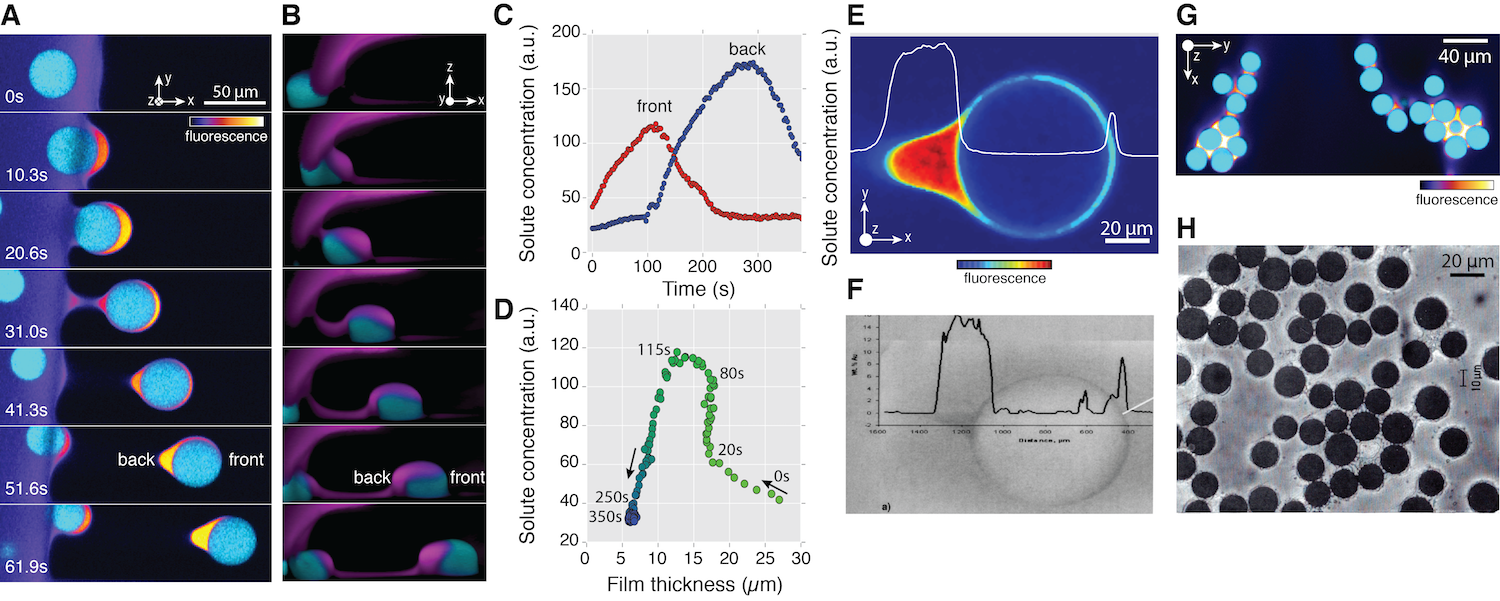}
\caption{Dynamics of premelted films around the droplets and impact on the solidification microstructure. A: Typical interaction between a droplet and the solidification front, that shows the accumulation of solute and its redistribution around the droplet. Horizontal cross-section. Sample velocity: 2~$\mu m/s$, temperature gradient: $5^\circ C/mm$. B: Same sequence, but in 3D, and in vertical cross-section. The meniscus formed by the solidification front can clearly be observed, as well as the premelted film dynamics. Most of the solute is eventually pushed below the crystal. C: Time-lapse evolution of solute concentration ahead and behind the droplet. D: Time-lapse evolution of solute concentration vs. film thickness. E: Solute-rich, comet-tail shaped region behind the droplet and corresponding solute concentration profile across the droplet in the frozen region. A similar solute segregation has been found in metal alloys (F), such as gold segregated around a porosity~\cite{Catalina2004}. The plot shows the gold concentration across the picture. A solute-rich, comet-tail shaped region is observed behind the pore. Reprinted by permission from Springer Nature: Metall. Mater. Trans. A \cite{Catalina2004}, copyright (2004) G: Agglomerates of droplets. The solute-rich regions around the droplets are the last regions to solidify. H: An analogous behaviour is observed in metals, where this segregation can lead to the formation of a secondary phase around objects, such as the $\theta (CuAl_2)$ phase in an Al-4.5wt.\% Cu alloy formed around alumina fibers. Micrograph courtesy of Pavel Bystricky and Andreas Mortensen~\cite{Bystricky1997}.  \label{fig:figure4}}
\end{figure}

\section*{Supplementary materials}

Fig S1--S6\\
Movie S1\\
References \textit{(30--31)}

\nocite{Marcellini2016}
\nocite{Schindelin2012a}

\end{document}